\documentclass[1p,times]{elsarticle}
\usepackage{graphicx}
\usepackage{color}
\usepackage{amssymb}
\usepackage{amsmath}
\usepackage{xspace}
\usepackage{multirow}   
\usepackage{booktabs}   
\usepackage{longtable}
\usepackage{xcolor}

\usepackage{hyperref}
\hypersetup{colorlinks=true, linkcolor=black, citecolor=black, urlcolor=black}
\usepackage{lineno}

\biboptions{comma, square, sort&compress}
\journal{Journal of Magnetism and Magnetic Materials} 
\begin{document}
\begin{frontmatter}



\ead{thomas.schrefl@donau-uni.ac.at}
\cortext[cor1]{Corresponding author} 

\title{Magnetostatics and micromagnetics with physics informed neural networks}
\author[a,b]{Alexander Kovacs}
\author[c,d,e]{Lukas Exl}
\author[a,b]{Alexander Kornell}
\author[a,b]{Johann Fischbacher}
\author[a,b]{Markus Hovorka}
\author[a,b]{Markus Gusenbauer}
\author[b]{Leoni Breth}
\author[b]{Harald Oezelt}
\author[f]{Dirk Praetorius}
\author[g,e]{Dieter Suess}
\author[a,b,e]{Thomas Schrefl\corref{cor1}}

\address[a]{Christian Doppler Laboratory for Magnet design through physics informed machine learning, Viktor Kaplan-Stra{\ss}e 2E, 2700 Wiener Neustadt, Austria}
\address[b]{Department for Integrated Sensor Systems, Danube University Krems, Viktor Kaplan-Stra{\ss}e 2E, 2700 Wiener Neustadt, Austria}
\address[c]{Department of Mathematics, University of Vienna, Oskar-Morgenstern-Platz 1, 1090 Vienna, Austria}
\address[d]{Wolfgang Pauli Institute, Oskar-Morgenstern-Platz 1, 1090 Vienna, Austria}
\address[e]{Research Platform MMM Mathematics-Magnetism-Materials, Oskar-Morgenstern-Platz 1, 1090 Vienna, Austria}
\address[f]{Institute of Analysis and Scientific Computing, TU Wien, Wiedner Hauptstra{\ss}e 8-10, 1040 Vienna, Austria}
\address[g]{Physics of Functional Materials, University of Vienna,  Währinger Stra{\ss}e 17, 1090 Vienna, Austria}


\begin{abstract}
Partial differential equations and variational problems can be solved with physics informed neural networks (PINNs). The unknown field is approximated with neural networks. Minimizing the residuals of the static Maxwell equation at collocation points or the magnetostatic energy, the weights of the neural network are adjusted so that the neural network solution approximates the magnetic vector potential. This way, the magnetic flux density for a given magnetization distribution can be estimated. With the magnetization as an additional unknown, inverse magnetostatic problems can be solved. Augmenting the magnetostatic energy with additional energy terms, micromagnetic problems can be solved. We demonstrate the use of physics informed neural networks for solving magnetostatic problems, computing the magnetization for inverse problems, and calculating the demagnetization curves for two-dimensional geometries.          
\end{abstract}

\begin{keyword}
magnetostatics \sep neural network \sep Ritz method \sep inverse problems
\end{keyword}
\end{frontmatter}


\section{\label{sec:introduction}Introduction}

The design and optimization of magnetic devices is linked with the ability to compute the magnetic flux, to solve inverse magnetostatic problems, and to calculate hysteresis loops of magnetic materials. 

Examples on the macroscopic scale are magnets in accelerators and electron storage rings \cite{halbach1985application}, magnetic write heads in magnetic recording \cite{bashir2012head}, and permanent magnet systems with a predefined stray field for example for sensor applications \cite{huber20173d}. In many applications, a magnetostatic field with predefined properties is sought: Accelerator magnets should produce fields that are either uniform or vary linearly in space. Magnetic recording heads that create fields with a high field gradient are essential to achieve high storage densities. In order to reach fields with certain properties, either the shape of the field source or the magnetization distribution within the field generating magnet \cite{huber20173d} or both can be optimized. For shape optimization, on/off methods \cite{takahashi2008advanced,abert2017fast}, in which space points are either magnetic or non-magnetic, or parameterized geometries \cite{kovacs2014numerical} have been used. Inverse problems, in which the optimal distribution of the magnetization is to be found, are efficiently solved with the adjoint method \cite{bruckner2017solving}.  

On a microscopic scale the computation of the magnetization inside a magnetic material is of importance. Solving for the magnetization as function of the external field gives the hysteresis loop. The magnetization distribution is the solution of Brown's micromagnetic equation \cite{brown1963micromagnetics}. Micromagnetics addresses the interplay between the local chemical composition, the microstructure of the material, and the hysteresis properties \cite{fischbacher2018micromagnetics, Exl2020}. Local material properties are reflected by the coefficients of the partial differential equation (PDE).

Traditionally, the numerical solution of (inverse) magnetostatic  and micromagnetic problems relies on the finite difference or finite element discretization of the underlying partial differential equations. For the fast estimation of magnetostatic fields in motors \cite{khan2019deep} or the magnetic response of magnetic sensor elements, neural networks \cite{khan2019deep,kovacs2019learning}  or kernel methods \cite{exl2020learning} have been applied. In order to train the machine learning models,  conventional numerical solvers are used to generate the training data by varying geometry, external loads, or time. This makes the numerical solution of the partial differential equations a preprocessing step. However, once the machine learning model is trained,  the magnetic field or the magnetization can be quickly estimated. Such models are useful for interactive design, optimization, or real time applications.    

An alternative way for solving partial differential equations numerically are physics informed neural networks \cite{raissi2019physics}. The loss function of physics informed neural networks, which is minimized during training, is directly computed from the governing partial differential equation. The loss function is either formed by the residuals at collocation points \cite{koryagin2019pydens}, the weighted residuals obtained by the Galerkin method \cite{kharazmi2019variational}, or the energy functional of an Euler-Lagrange differential equation \cite{e2018deep}. For training a physics informed neural network, there is no need to generate training data in advance. The input data for physics informed neural networks are points sampled in the problem domain. 

The loss function of physics informed neural networks can be augmented with the distance between the approximated solution of the partial differential equation and desired values of the solution at given points in space. Then, one or more coefficients of the partial differential equation can be included as unkowns during training. Solving inverse problems with physics informed neural networks may lead to a significant speed up as compared to conventional methods \cite{hennigh2020nvidia}.

\begin{figure}[!tb]
	\centering
	\includegraphics[scale=0.6]{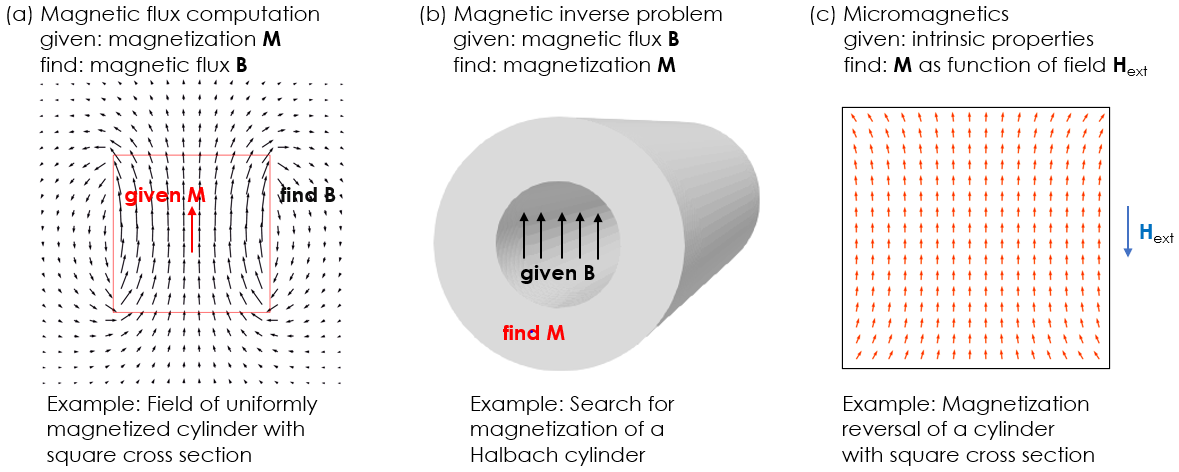}
	\caption{\label{fig:applications}Applications for physics informed neural networks addressed in this work. (a) Magnetostatic field computation: Estimation of the magnetic flux of a permanent magnet. (b) Magnetostatic inverse problem: Estimation of the magnetization distribution for a Halbach cylinder. (c) Micromagnetics: Computation of the hysteresis loop of a hard magnetic particle. To demonstrate the use of physics informed neural networks in magnetostatics and micromagentics for two-dimensional problems. The magnets are infinitely extended in one direction. This is schematically shown in (b). The magnets in (a) and (b) are infinitely extended in the direction normal to the drawing plane.}
\end{figure}

In this work, we demonstrate the use of physics informed neural networks for inverse magnetostatic and micromagnetic problems. We first show how physics informed neural networks can be used to solve the forward magnetostatic problem. We use a set of dense neural networks \cite{e2018deep} to approximate the unkown magnetic vector potential. During training, the residuals related to the partial differential equation and boundary and interface conditions are minimized by adjusting the weights of the network. For solving inverse problems, a second set of neural networks approximates the unknown magnetization. An additional condition is added to the sum of residuals that penalizes the difference between the desired field and the current numerical estimate. 

Alternatively, we can compute the magnetic flux density created from a given magnetization distribution by minimizing Brown's upper limit for the magnetostatic energy \cite{brown1963micromagnetics}. During training, the magnetostatic energy is minimized by adjusting the weights of the network. This approach is similar to the Ritz method for computing the magnetic field: However, instead of finite element basis functions \cite{guancial1977three}, we use dense neural networks \cite{e2018deep} to approximate the unkowns. Again we can introduce the magnetization as additional unknown and approximate it with a set of neural networks. We can minimize the total micromagnetic energy, which is the sum of the magnetostatic energy, the ferromagnetic exchange energy, the magneto-crystalline anisotropy energy, and the exchange energy by simultaneously adjusting the weights in the neural networks for the magnetic vector potential and the magnetization. This joint minimization of the energy with respect to the magnetic vector potential and the magnetization for the numerical solution of micromagnetic problems was suggested by Asselin and Thiele \cite{asselin1986field} and applied in finite element micromagnetics for soft magnetic elements \cite{fredkin1987numerical} and permanent magnets \cite{schrefl1994two}. 

Figure~\ref{fig:applications} show the problems addressed with physics informed neural networks within this work. Whenever possible we will focus on simple problems for which analytical solutions are known. In a magnetostatic problem we compute the magnetic flux density for a given magnetization distribution. As we will compute the magnetic flux density of a uniformly magnetized particle \cite{gronefeld1989calculation} with physics informed neural networks. In magnetostatic innverse problems we search for the magnetization when the magnetic flux density is given. As an example we will compute the magnetization distribution in the ring of a Halbach cylinder \cite{halbach1980design}. In micromagnetics we search for the magnetization distribution as a function of the applied field. Please note that for this problem we also have to take care of hysteresis: The magnetic states depends on the history of the applied field. Here we compute the coercive field of a Nd$_2$Fe$_{14}$B particle \cite{schrefl1994two}. We will restrict ourselves to two-dimensional problems. In the past, two-dimensional micromagnetic simulations of permanent magnets were found to give reasonable lower bounds for the coercive field \cite{schrefl1994two}. The comparison of experimental data with two-dimensional micromagnetic results showed excellent agreement for the remanent magnetization and the coercive field \cite{schrefl1994remanence}.  

The paper is organized as follows. We will first introduce the governing partial differential equations of magnetostatics and the micromagnetic energy functional. Then we will show how the solutions can be approximated with neural networks and we will define the loss functions associated with each problem. Finially we will discuss the numerical results.

\section{Micromagnetic background}
\subsection{Two-dimensional magnetostatics}
We are interested in computing the magnetic flux density, $\mathbf B$, or the magnetic field, $\mathbf H$, for a given magnetization distribution, $\mathbf M$.
In magnetostatics we have no time dependent quantities. In the
presence of a stationary current Maxwell’s equations reduce to \cite{steele2012numerical} 
\begin{align}
\label{eq:maxwellH}
\mathbf{\nabla} \times \mathbf{H} & = \mathbf{j},\\
\label{eq:maxwellB}
\mathbf{\nabla} \cdot \mathbf{B} & = 0.
\end{align}
Here $\mathbf j$ is the current density. The charge density fulfills $\mathbf \nabla \cdot \mathbf j = 0$ which expresses the conservation of electric charge. On a macroscopic length scale the relation between the magnetic induction
and the magnetic field is expressed by
\begin{align}
	\label{eq:material}
	\mathbf B = \mu \mathbf H,
\end{align}
where $\mu$ is the permeability of the material. Equation (\ref{eq:material}) is used in magnetostatic field solvers \cite{steele2012numerical} for the design of magnetic circuits. In these simulations, the permeability describes the response of the material to the magnetic field. The influence of the material can also be expressed by its magnetization distribution $\mathbf{M}(\mathbf x)$. Then we use 
\begin{align}
	\label{eq:BisHplusM}
	\mathbf B = \mu_0(\mathbf H + \mathbf M)
\end{align}
instead of (\ref{eq:material}). Here $\mu_0$ is the permeability of vacuum. For now, let us assume $\mathbf j = 0$. Taking the curl of (\ref{eq:BisHplusM}) and plugging into (\ref{eq:maxwellH}) gives
\begin{align}
	\label{eq:rotB}
	\mathbf \nabla \times \mathbf B = \mu_0 \mathbf \nabla \times \mathbf M.
\end{align}
In numerical calculations, the
constraint (\ref{eq:maxwellB}) that $\mathbf{B}$ is solenoidal can be fulfilled by introducing a magnetic vector potential $\mathbf{\nabla} \times \mathbf{A} = \mathbf{B}$. Thus, we arrive at 
\begin{align}
	\label{eq:rotrotA}
	\mathbf \nabla \times \left( \mathbf \nabla \times \mathbf A \right) = \mu_0 \mathbf \nabla \times \mathbf M.
\end{align}
We now assume that the magnetic sources are infinitely extended in direction $x_3$ and that the  magnetization  $\mathbf{M}$ is translationally invariant in $x_3$. Then the $x_3$ component of the magnetic induction is constant. The problem reduces to two dimensions \cite{asselin1986field} with $\mathbf{M} = \mathbf{M}(x_1,x_2)$, $\mathbf{B} = \mathbf{B}(x_1,x_2)$, and $A = A_{x_3}(x_1,x_2)$. For simplicity, we simply write $A$ for the $x_3$-component of the magnetic vector potential. The components of the magnetic flux density are  \cite{halbach1980design} 
\begin{align}
	\label{eq:2d}
	B_{x_1} = \frac{\partial A}{\partial x_2}, \;
	B_{x_2} = -\frac{\partial A}{\partial x_1}.
\end{align}

Using the vector identity $\mathbf \nabla \times \left( \mathbf \nabla \times \mathbf A \right) = \mathbf \nabla ( \mathbf \nabla \cdot \mathbf A) - \nabla^2 \mathbf A$ and translational invariance in $x_3$ direction, we rewrite (\ref{eq:rotrotA}) as
\begin{align}
	\label{eq:rotrotA2d}
	\frac{\partial^2 A}{\partial x_1^2} + \frac{\partial^2 A}{\partial x_2^2} = \mu_0 \left( \frac{\partial M_{x_1}}{\partial x_2} - \frac{\partial M_{x_2}}{\partial x_1} \right).
\end{align}
Equation (\ref{eq:rotrotA2d}) has to be fulfilled inside the magnetic material where $|\mathbf{M}| > 0$. Outside of the magnet, the magnetic vector potential solves the Laplace equation
\begin{align}
	\label{eq:rotrotA2dout}
	\frac{\partial^2 A}{\partial x_1^2} + \frac{\partial^2 A}{\partial x_2^2} = 0.
\end{align}
At the surface of the magnetic body with normal vector $\mathbf n$, the normal component of the magnetic flux density and the tangential component of the magnetic field are continuous  
\begin{align}
\label{eq:flux2d}
\left( \mathbf B^\mathrm{(in)} - \mathbf B^\mathrm{(out)}\right) \cdot \mathbf n &= 0,\\
\left( \mathbf H^\mathrm{(in)} - \mathbf H^\mathrm{(out)} \right) \times \mathbf n &= 0.
\end{align}
In terms of the magnetic vector potential and the magnetization, the interface conditions are
\begin{align}
	\label{eq:Bn2d}
	\left( \frac{\partial A^\mathrm{(in)}}{\partial x_2} - \frac{\partial A^\mathrm{(out)}}{\partial x_2} \right) n_{x_1} + \left( \frac{\partial A^\mathrm{(out)}}{\partial x_1} - \frac{\partial A^\mathrm{(in)}}{\partial x_1} \right) n_{x_2}&= 0,\\
	\label{eq:Ht2d}
	\left( \frac{\partial A^\mathrm{(in)}}{\partial x_2} - \frac{\partial A^\mathrm{(out)}}{\partial x_2} - \mu_0 M_{x_1} \right) n_{x_2} -  \left( \frac{\partial A^\mathrm{(out)}}{\partial x_1} - \frac{\partial A^\mathrm{(in)}}{\partial x_1} - \mu_0  M_{x_2} \right) n_{x_1} &= 0.
\end{align}
At infinity the magnetic flux density approaches zero.

Alternatively, we can minimize a sharp upper bound of the magnetostatic energy. Brown \cite{brown1964some} suggested a functional to be used as upper bound for the magnetostatic energy. The magnetostatic energy $E_\mathrm{m}$ due to the magnetization $\mathbf{M}(\mathbf{x})$ can be bounded by a functional \cite{schrefl1994two}
\begin{align}
	\label{eq:brown}
	E_\mathrm{m} \le W_\mathrm{m}(\mathbf{B}') = \frac{1}{2\mu_0} \int \left(\mathbf{B}'-\mu_0\mathbf{M}\right)^2 \mathrm{d}^3x.
\end{align}
The functional $W_\mathrm{m}(\mathbf{B}')$ if minimized subject to the constraint $\mathbf{\nabla} \cdot \mathbf{B}' = 0$, makes $\mathbf{B}'$ equal to the magnetic induction $\mathbf{B}$ created by the magnetization $\mathbf{M}$. Again, we can introduce a magnetic vector potential to make  $\mathbf{B}'$ solenoidal. The Euler-Lagrange equation of (\ref{eq:brown}) with respect to the magnetic vector potential gives the partial differential equation (\ref{eq:rotrotA}) \cite{asselin1986field}.

Assuming translational symmetry in $x_3$ direction, we obtain
\begin{align}
	\label{eq:brown2d}
	W_\mathrm{m}(A') = \frac{1}{2\mu_0} \int \left( \left(\frac{\partial A'}{\partial x_1}\right)^2 + \left(\frac{\partial A'}{\partial x_2}\right)^2 + 2 \mu_0 \left(M_{x_2}\frac{\partial A'}{\partial x_1} - M_{x_1}\frac{\partial A'}{\partial x_2}\right) + \mu_0^2\left(M_{x_1}^2 + M_{x_2}^2 \right)\right) \mathrm{d}^2x.
\end{align}
If minimized with respect to $A'$, the functional (\ref{eq:brown2d}) gives the magnetostatic energy. Please note that the last term in (\ref{eq:brown2d}) adds a constant offset to the magnetostatic energy and thus may be dropped.  

The integral in (\ref{eq:brown2d}) is over the entire space. Thus, when evaluating (\ref{eq:brown2d}), we have to integrate over the magnet, where $M_{x_1} \ne 0$ or $M_{x_2} \ne 0$, and over a large region outside the magnetic material. The distance of the outer 
boundary to the center of the magnet  should be at least five times the distance of the center of the magnet to its most remote outer surface \cite{chen1997review}. 

\subsection{Two-dimensional micromagnetics}

In micromagnetics, we want to compute the local distribution of the magnetization as function of the magnetic field. This is the response of the system to (an external) field. For a given value of the external field $\mathbf H_\mathrm{ext}$, the magnetization distribution  $\mathbf{M} = \mathbf{M}(x_1,x_2)$ can be derived from the minimization of the total Gibbs free energy. Upon minimization, the constraint $\left|\mathbf M\right|=M_\mathrm{s}$ has to be fulfilled. The spontaneous magnetization $M_\mathrm{s}$ of a material depends on temperature, but is independent of the external field \cite{brown1963micromagnetics}. The hysteresis loop follows from the path formed by subsequently following local minima in an energy landscape progressively changed by a varying external field \cite{kinderlehrer1994simulation}.

The Gibbs free energy $E$ is the sum of the magnetostatic energy $E_\mathrm{m}$, the Zeeman energy of the magnetization in an external field, the magneto-crystalline anisotropy energy, and the ferromagnetic exchange energy \cite{brown1963micromagnetics}. For an efficient numerical scheme we follow Asselin and Thiele \cite{asselin1986field} and replace $E_\mathrm{m}$ with $W_\mathrm{m}$. We define an upper bound of the total energy
\begin{align}
	\label{eq:mumag}
	W(\mathbf M, A') = W_\mathrm{m}(\mathbf M, A') &+ \int_{V^\mathrm{(in)}} \left\{-\mu_0 \mathbf M \cdot \mathbf H_\mathrm{ext} + K_1 \sin^2\alpha + K_2 \sin^4\alpha \right. \nonumber \\ & \left. + \frac{C}{M_\mathrm{s}^2}\left[\left(\nabla M_{x_1} \right)^2 +\left(\nabla M_{x_2} \right)^2 \right] \right\}\mathrm{d}^2x.
\end{align}  
The local minima of the auxiliary functional $W$ are in one-to-one correspondence with those of the total Gibbs free energy $E$.  Here $K_1$, $K_2$ are the anisotropy constants, $\alpha$ is the angle between the magnetization $\mathbf M$ and the  anisotropy direction, and $C$ is the exchange constant. The second term on the right hand side of (\ref{eq:mumag}) is over the volume of the magnet $V^\mathrm{(in)}$ where $|\mathbf M| > 0$.  The right hand side of (\ref{eq:mumag}) contains $W_m(\mathbf M, A')$, which is an upper bound for the magnetostatic energy. Minimization of the right hand side of (\ref{eq:mumag}) with respect to $A'$ makes $W_m(\mathbf M, A')$ equal to the magnetostatic energy $E_\mathrm{m} = E_\mathrm{m}(\mathbf M)$.   

\section{Physics informed neural networks}
\subsection{Collocation based magnetostatics}
For creating the neural network approximation, we follow the approach of Niakia and co-workers \cite{niaki2020physics} and introduce distinct neural networks for the different regions of the problem domain. The two neural networks approximate the vector potentials inside the magnetic domain, $A^\mathrm{(in)}$, and outside the magnetic domain,  $A^\mathrm{(out)}$, respectively. The interface conditions (\ref{eq:Bn2d}) and (\ref{eq:Ht2d}) are incorporated into the loss function. The inputs of the neural networks are points in the two-dimensional problem domain $(x_1,x_2)$, the output of the neural network is the approximation of the magnetic vector potential:
\begin{align}
  \label{eq:nnAin}
  A^\mathrm{(in)}_\mathrm{approx} &= \mathcal{N}_{A^\mathrm{(in)}}(x_1,x_2,\mathbf w_{A^\mathrm{(in)}}), \\
  \label{eq:nnAout}
  A^\mathrm{(out)}_\mathrm{approx} &= \mathcal{N}_{A^\mathrm{(out)}}(x_1,x_2,\mathbf w_{A^\mathrm{(out)}}).
\end{align} 
The vectors $\mathbf w_{A^\mathrm{(in)}}$ and  $\mathbf w_{A^\mathrm{(out)}}$ represent the weights and biases of each network. The location of a point in the problem domain is given by the vector $\mathbf x = (x_1,x_2)$. The weights and biases are the learnable parameters of the networks which are determined during training of the networks by minimizing the sum of the squared residuals at collocation points. During training of the neural networks (\ref{eq:nnAin}) and (\ref{eq:nnAout}) the weights and biases are adjusted so that $A^\mathrm{(in)}_\mathrm{approx}$ is an approximate solution of equation (\ref{eq:rotrotA2d}), $A^\mathrm{(out)}_\mathrm{approx}$ is an approximate solution of equation (\ref{eq:rotrotA2dout}), and both fulfill the interface conditions (\ref{eq:Bn2d}) and (\ref{eq:Ht2d}). The magnetic flux should decay to zero as $|\mathbf x|$ approaches infinity. To account for this condition we expand the problem domain up to a certain distance outside the magnetic body and force $\mathbf B = 0$ at the boundary of the truncated problem domain. The two networks are trained simultaneously. The loss function for the joint training of the two networks is the following sum  
\begin{align}
	\label{eq:loss}
	L_\mathrm{collocation} = L_{A^\mathrm{(in)}} + L_{A^\mathrm{(out)}} + L_{\mathbf B^\mathrm{(out)}} + L_{B_\mathrm{n}} + L_{H_\mathrm{t}}. 
\end{align} 
We define the following indicator functions or binary masks
\begin{align}
  in(\mathbf x)   &= 1 \; \mathrm{for}\; \mathbf x  \; \mathrm{inside}\; V^\mathrm{in}, \\
  in(\mathbf x)   &= 0 \; \mathrm{otherwise}, \\
  bnd(\mathbf x)  &= 1 \; \mathrm{for}\; \mathbf x \; \mathrm{close\, to\, the\, surface\, of}\; V^\mathrm{in}, \\
  bnd(\mathbf x)  &= 0 \; \mathrm{otherwise}, \\
  inf(\mathbf x)  &= 1 \; \mathrm{for}\; \mathbf x \; \mathrm{close\, to\, the\, outer\, boundary\, of\, the\, problem\, domain,\, and} \\
  inf(\mathbf x)  &= 0 \; \mathrm{otherwise}, 
\end{align} 
in order to track the location of the collocation points $\mathbf x_i$. 

The loss $L_{A^\mathrm{(in)}}$ is the mean squared sum of the residuals of ($\ref{eq:rotrotA2d}$)
\begin{align}
  L_{A^\mathrm{(in)}} = \frac{1}{n} \sum_{i=1}^n in(\mathbf x_i)&\left( \left.\frac{\partial^2 {A_\mathrm{approx}^\mathrm{(in)}(\mathbf x)}}{\partial x_1^2}\right|_{\mathbf x = \mathbf x_i} + \left.\frac{\partial^2 {A_\mathrm{approx}^\mathrm{(in)}(\mathbf x)}}{\partial x_2^2}\right|_{\mathbf x = \mathbf x_i} \right. \\ \nonumber &\left. - \mu_0 \left(\left. \frac{\partial M_{x_1}(\mathbf x)}{\partial x_2}\right|_{\mathbf x = \mathbf x_i} - \left.\frac{\partial M_{x_2}(\mathbf x)}{\partial x_1}\right|_{\mathbf x = \mathbf x_i} \right)\right)^2.
\end{align}
Terms denoted with the subscript 'approx' such as $A_\mathrm{approx}^\mathrm{(in)}(\mathbf x)$ are the neural network approximations of the respective physical quantities and are continuous functions in space when the hyperbolic tangent is used as activation function for the hidden layers. The loss $L_{A^\mathrm{(out)}}$ is the mean squared sum of the residuals of ($\ref{eq:rotrotA2dout}$)
\begin{align}
	L_{A^\mathrm{(out)}} = \frac{1}{n} \sum_{i=1}^n \left(1-in(\mathbf x_i)\right)\left( \left.\frac{\partial^2 {A_\mathrm{approx}^\mathrm{(out)}(\mathbf x)}}{\partial x_1^2}\right|_{\mathbf x = \mathbf x_i} + \left.\frac{\partial^2 {A_\mathrm{approx}^\mathrm{(out)}(\mathbf x)}}{\partial x_2^2}\right|_{\mathbf x = \mathbf x_i} \right)^2.
\end{align}
Plugging $A^\mathrm{(in)}_\mathrm{approx}$ and $A^\mathrm{(out)}_\mathrm{approx}$ into equation (\ref{eq:2d}), we get the neural network approximations of the magnetic flux density inside $\mathbf B^\mathrm{(in)}_\mathrm{approx}$ and outside the magnet $\mathbf B^\mathrm{(out)}_\mathrm{approx}$. Upon minimization, the loss $L_{\mathbf B^\mathrm{(out)}}$ makes the magnetic flux density vanish at the outer boundary of the problem domain:
\begin{align}
	L_{\mathbf B^\mathrm{(out)}} = \frac{1}{n} \sum_{i=1}^n inf(\mathbf x_i) \left( \left(B_{x_1,\mathrm{approx}}^\mathrm{(out)}(\mathbf x_i) \right)^2 + \left(B_{x_2,\mathrm{approx}}^\mathrm{(out)}(\mathbf x_i) \right)^2 \right).
\end{align}
Similarly to the neural network approximation of the magnetic flux density, we can derive the neural network approximation of the magnetic field inside $\mathbf H^\mathrm{(in)}_\mathrm{approx}$ and outside the magnet $\mathbf H^\mathrm{(out)}_\mathrm{approx}$ from the neural network approximations of the magnetic vector potential. The following two loss functions account for the continuity of the normal component of the magnetic flux density and the continuity of the tangential component of the magnetic field at the surface of the magnet:
\begin{align}
	L_{B_\mathrm{n}} &= \frac{1}{n} \sum_{i=1}^n bnd(\mathbf x_i) \left( \left( \mathbf B^\mathrm{(in)}_\mathrm{approx}(\mathbf x_i) - \mathbf B^\mathrm{(out)}_\mathrm{approx}(\mathbf x_i)\right) \cdot \mathbf n(\mathbf x_1)  \right)^2, \\
   	L_{H_\mathrm{t}} &= \frac{1}{n} \sum_{i=1}^n bnd(\mathbf x_i) \left( \left( \mathbf H^\mathrm{(in)}_\mathrm{approx}(\mathbf x_i) - \mathbf H^\mathrm{(out)}_\mathrm{approx}(\mathbf x_i) \right) \times \mathbf n(\mathbf x_i) \right)^2.
\end{align}

During training of the neural network, the loss is minimized with the stochastic gradient descent algorithm. Input data for training are quasi-randomly sampled points from the problem domain. The training set contains  $N$ points, which are quasi-randomly sampled with a Sobol sequence \cite{sobol1976uniformly}. For training, the points are combined into batches. In each iteration step, the stochastic gradient descent method adjusts the weights of the neural network according to the samples of one batch. The batch size $n < N$ is the number of points which are used to evaluate the loss function (\ref{eq:loss}). The batch size needs to be large enough so that each batch contains points in the magnet, outside the magnet, close to the magnet's surface and close to the outer boundary \cite{haghighat2021sciann}. During optimization, the algorithm passes several times through the complete training set. 

\begin{figure}[!tb]
	\centering
	\includegraphics[scale=0.35]{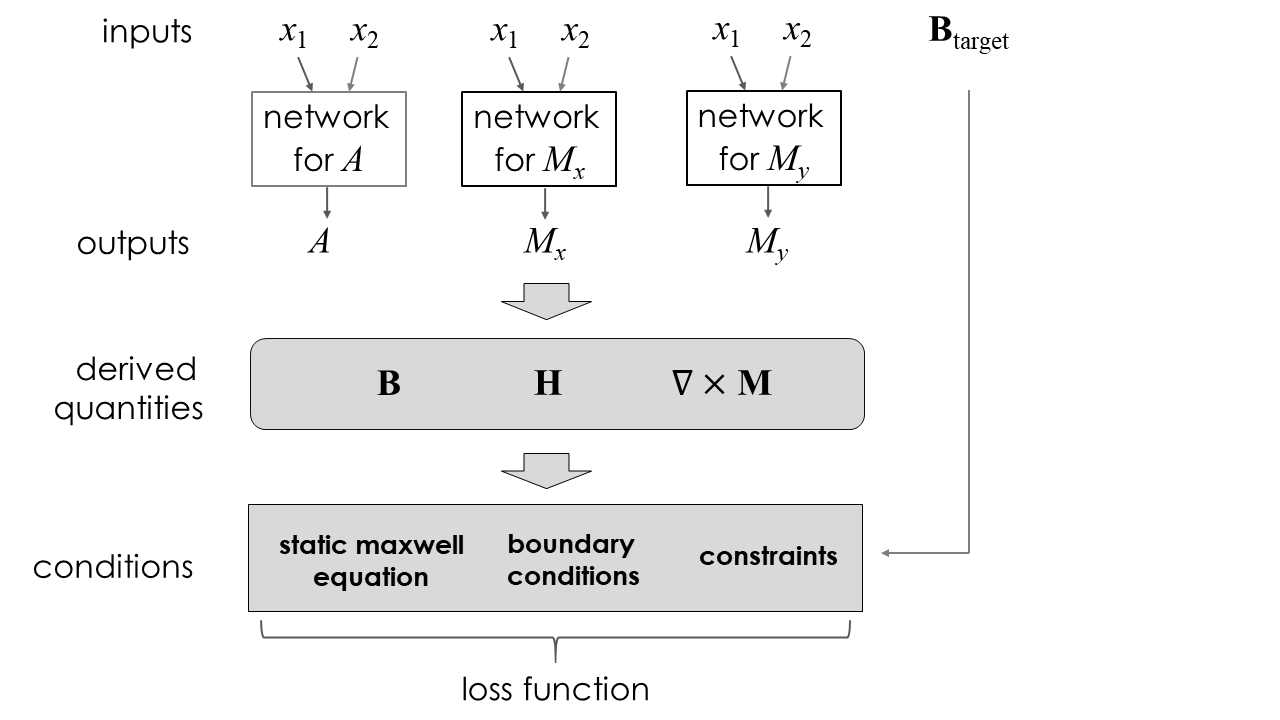}
	\caption{\label{fig:schematics1}Schematics of a physics informed neural network for solving magnetostatic inverse problems. Dense neural networks approximate the magnetic vector potential $A$ and the magnetization components $M_{x_1}$ and $M_{x_2}$. Inputs for the networks are the location of the points in the problem domain.
	From the network outputs the terms in the underlying physical equations are derived. The loss function for training the networks is the sum of squares of the residuals associated with the static Maxwell equation, boundary conditions, and constraints, which involve target values for the magnetic flux density.}
\end{figure}
\subsection{Inverse magnetostatic problems}
\label{sec:inverse}
For inverse magnetostatic problems, we introduce additional neural networks 
\begin{align}
  \label{eq:nnMx}
M_{x_1,\mathrm{approx}} &= \mathcal{N}_{M_{x_1}}(x_1,x_2,\mathbf w_{M_{x_1}}), \\
 \label{eq:nnMy}
M_{x_2,\mathrm{approx}} &= \mathcal{N}_{M_{x_2}}(x_1,x_2,\mathbf w_{M_{x_2}}), 
\end{align}
that approximate the unknown magnetization distribution. The loss function for training is augmented with loss functions for the constraint
$\left| \mathbf M_\mathrm{approx} \right|=M_\mathrm{s}$. The set of governing partial differental equations, boundary conditions, and constraints have to reflect the  target magnetic flux density 
$\mathbf B_\mathrm{target}$ of the problem. 
One prominent magnetostatic inverse problem is the Halbach cylinder (see Figure~\ref{fig:applications}b). We want to find the orientation of the magnetization in a long magnetic cylinder that generates a uniform vertical field in a cylindrical cavity and is zero outside the magnetic system. For simplicity we consider the cylinder to be infinitely extended and make use of translational symmetry along the cylinder axis. For the Halbach cylinder the governing equations are (\ref{eq:rotrotA2d}) and the interface conditions 
\begin{align}
	\left( \mathbf B^\mathrm{(in)} - \mathbf B_\mathrm{target}^\mathrm{(cavity)}\right) \cdot \mathbf n &= 0 \;\mathrm{at\,the\,inner\,surface\,of\,the\,cylinder},\\
	\left( \mathbf H^\mathrm{(in)} - \mathbf H_\mathrm{target}^\mathrm{(cavity)} \right) \times \mathbf n &= 0 \;\mathrm{at\,the\,inner\,surface\,of\,the\,cylinder},\\
	\mathbf B^\mathrm{(in)}\cdot \mathbf n &= 0\;\mathrm{at\,the\,outer\,surface\,of\,the\,cylinder},\\
    \mathbf H^\mathrm{(in)} \times \mathbf n &= 0 \;\mathrm{at\,the\,outer\,surface\,of\,the\,cylinder}.
\end{align}
The magnetostatic inverse problem for the Halbach cylinder leads to the following loss function
\begin{align}
	\label{eq:lossinv}
	L_\mathrm{halbach} = L_{\mathbf M} + L_{A^\mathrm{(in)}} + L_{B_\mathrm{n},1} + L_{H_\mathrm{t},1} + L_{B_\mathrm{n},2} + L_{H_\mathrm{t},2}. 
\end{align} 
We introduce the indicator functions $bnd_1(\mathbf x)$ and $bnd_2(\mathbf x)$ to select training points close to the inner and outer surface of the cylinder, respectively.
The individual loss functions are
\begin{align}
L_{\mathbf M} &= \frac{1}{n} \sum_{i=1}^n in(\mathbf x_i) \left( \sqrt{M_{x_1,\mathrm{approx}}^2+M_{x_2,\mathrm{approx}}^2}-1 \right)^2, 
\end{align}
\begin{align}
  L_{A^\mathrm{(in)}} = \frac{1}{n} \sum_{i=1}^n in(\mathbf x_i)&\left( \left.\frac{\partial^2 {A_\mathrm{approx}^\mathrm{(in)}(\mathbf x)}}{\partial x_1^2}\right|_{\mathbf x = \mathbf x_i} + \left.\frac{\partial^2 {A_\mathrm{approx}^\mathrm{(in)}(\mathbf x)}}{\partial x_2^2}\right|_{\mathbf x = \mathbf x_i} \right. \\ \nonumber &\left.  - \mu_0 \left( \left.\frac{\partial M_{x_1,\mathrm{approx}}(\mathbf x)}{\partial x_2}\right|_{\mathbf x = \mathbf x_i} - \left.\frac{\partial M_{x_2,\mathrm{approx}}(\mathbf x)}{\partial x_1}\right|_{\mathbf x = \mathbf x_i} \right)\right)^2,
\end{align}
\begin{align}
  	L_{B_\mathrm{n},1} &= \frac{1}{n} \sum_{i=1}^n bnd_1(\mathbf x_i) \left( \left( \mathbf B^\mathrm{(in)}_\mathrm{approx}(\mathbf x_i) - \mathbf B^\mathrm{(cavity)}_\mathrm{target}(\mathbf x_i)\right) \cdot \mathbf n(\mathbf x_i)  \right)^2, \\
  L_{H_\mathrm{t},1} &= \frac{1}{n} \sum_{i=1}^n bnd_1(\mathbf x_i) \left( \left( \mathbf H^\mathrm{(in)}_\mathrm{approx}(\mathbf x_i) - \mathbf H^\mathrm{(cavity)}_\mathrm{target}(\mathbf x_i) \right) \times \mathbf n(\mathbf x_i) \right)^2,\\
  	L_{B_\mathrm{n},2} &= \frac{1}{n} \sum_{i=1}^n bnd_2(\mathbf x_i) \left(  \mathbf B^\mathrm{(in)}_\mathrm{approx}(\mathbf x_i)  \cdot \mathbf n(\mathbf x_1)  \right)^2, \\
L_{H_\mathrm{t},2} &= \frac{1}{n} \sum_{i=1}^n bnd_2(\mathbf x_i) \left( \mathbf H^\mathrm{(in)}_\mathrm{approx}(\mathbf x_i)  \times \mathbf n(\mathbf x_i) \right)^2.
\end{align}


The schematics of a physics informed neural network for the solution of magnetostatic inverse problems is given in Figure~\ref{fig:schematics1}.

In inverse modeling regularization terms ensure the smoothness of the solution \cite{abert2017fast}. However, we found that such an explicit regularization is not required for solving inverse magnetostatic problems with physics informed neural networks.
We use neural networks with the hyperbolic tangent as activation function for the hidden layers. The neural network approximations of the magnetization components (\ref{eq:nnMx}) and (\ref{eq:nnMy}) are smooth functions. In addition, our networks are simple containing only a few hidden layers.  We speculate that the architecture takes the role of the regularization term commonly used in inverse modeling.

\subsection{Ritz based micromagnetics}
\label{sec:mumag}
Alternatively, we can apply the deep Ritz method as introduced by E and Yu \cite{e2018deep}, in order to solve the magnetostatic field problem. Then the loss function is the magnetostatic energy functional (\ref{eq:brown2d}). With the deep Ritz method for solving magnetostatic problems we use a single neural network 
\begin{align}
	\label{eq:nnA}
	A_\mathrm{approx} &= \mathcal{N}_{A}(x_1,x_2,\mathbf w_{A})
\end{align}
for the approximation of the magnetic vector potential. The weights and biases, which are represented by the vector $\mathbf w_{A}$, are determined during training of the network by minimizing the functional (\ref{eq:brown2d}). In order to evaluate this integral, we apply Monte-Carlo integration with quasi-randomly sampled points $\mathbf x_i$. The use of quasi-random points for Monte-Carlo integration improves convergence \cite{caflisch1998monte} since clumps of points that occur for random sampling can be avoided. Similarly, Hennig and co-workers \cite{hennigh2020nvidia} apply quasi-Monte Carlo integration to evaluate the integrals occurring during the solution of partial differential equations with physics informed neural networks. The loss function for training of the neural network is 
\begin{align}
	\label{eq:Lritz}
	L_\mathrm{mag} = \frac{V_\mathrm{dom}}{n} \frac{1}{2} \sum_{i=1}^n & \left( \left(\left.\frac{\partial A_\mathrm{approx}(\mathbf x)}{\partial x_1}\right|_{\mathbf x = \mathbf x_i}\right)^2 + \left(\left.\frac{\partial A_\mathrm{approx}(\mathbf x)}{\partial x_2}\right|_{\mathbf x = \mathbf x_i}\right)^2 \right. \nonumber \\
	&\left. + 2 \mu_0\, in(\mathbf x_i) \left(M_{x_2}(\mathbf x_i)\left.\frac{\partial A_\mathrm{approx}(\mathbf x)}{\partial x_1}\right|_{\mathbf x = \mathbf x_i} - M_{x_1}(\mathbf x_i)\left.\frac{\partial A_\mathrm{approx}(\mathbf x)}{\partial x_2}\right|_{\mathbf x = \mathbf x_i}\right) \right)
\end{align}
Here $V_\mathrm{dom}$ is the area of the problem domain. 

\begin{figure}[!tb]
	\centering
	\includegraphics[scale=0.35]{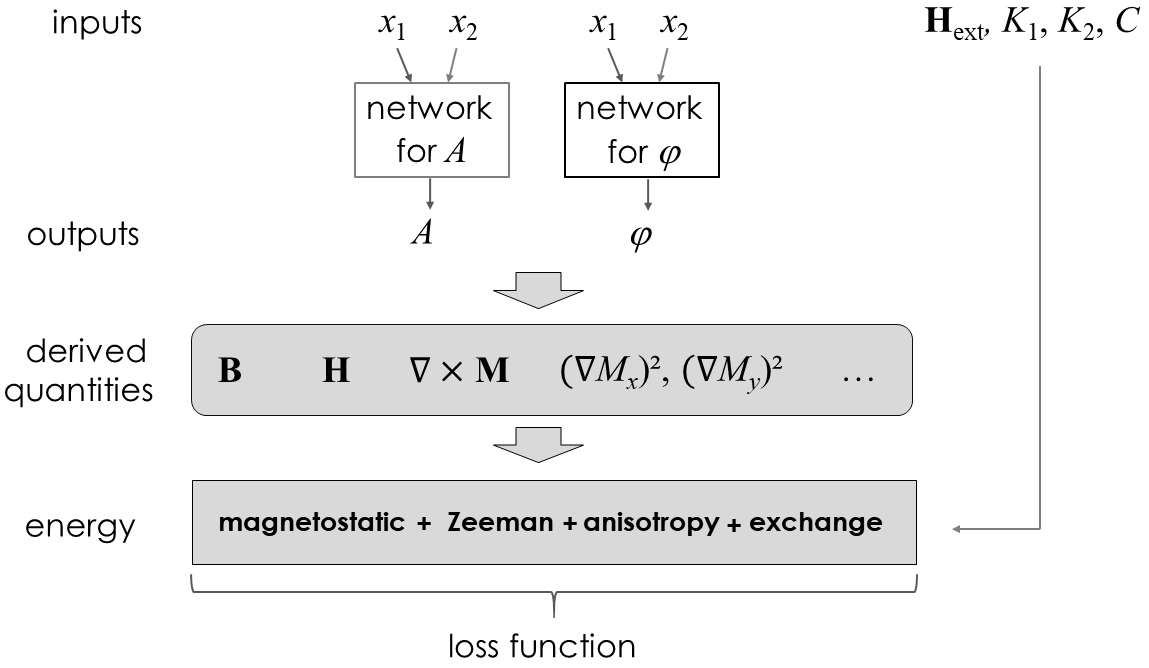}
	\caption{\label{fig:schematics2}Schematics of a physics informed neural network for solving micromagnetic problems. Dense neural networks approximate the magnetic vector potential $A$ and the magnetization angle $\varphi$. Inputs for the networks are points in the problem domain. From the network outputs the terms in the in the micromagnetic energy densities are derived. The loss function for training the networks is the total Gibbs free energy, which is evaluated by quasi-Monte-Carlo integration.}
\end{figure}

In micromagnetics we want to compute the magnetization $\mathbf M$. In addition to the neural network (\ref{eq:nnA}) we introduce an additional neural network 
\begin{align}
	\label{eq:nnphi}
	\varphi_\mathrm{approx} &= \mathcal{N}_{\varphi}(x_1,x_2,\mathbf w_{\alpha})
\end{align}
for the angle of the magnetization with respect to the $x_1$ axis. The neural network approximation of the magnetization components are given by
\begin{align}
	M_{x_1,\mathrm{approx}} &= M_\mathrm{s} \cos(\varphi_\mathrm{approx}),\\
	M_{x_2,\mathrm{approx}} &= M_\mathrm{s} \sin(\varphi_\mathrm{approx}).
\end{align}
The total loss function is the sum
\begin{align}
	L = L_\mathrm{mag} + L_\mathrm{zee} + L_\mathrm{ani} +L_\mathrm{ex}.
\end{align}
The summands represent the magnetostatic energy, the Zeeman energy, the anisotropy energy, and the exchange energy.  These energies are again evaluated with quasi-Monte-Carlo integration. The schematics of a physics informed neural network for micromagnetic simulations is shown in Figure~\ref{fig:schematics2}. In addition to $L_\mathrm{mag}$, the loss functions are
\begin{align}
  L_\mathrm{zee} &= \frac{\mu_0 V_\mathrm{dom}}{n} \sum_{i=1}^n in(\mathbf x_i) \left( -\mu_0 M_{x_1,\mathrm{approx}}(\mathbf x_i) H_\mathrm{ext,x_1} -\mu_0 M_{x_2,\mathrm{approx}}(\mathbf x_i) H_\mathrm{ext,x_2}\right), \\
  L_\mathrm{ani} &= \frac{\mu_0 V_\mathrm{dom}}{n} \sum_{i=1}^n in(\mathbf x_i) \left( K_1 \left(\frac{M_{x_1,\mathrm{approx}}(\mathbf x_i)}{M_\mathrm{s}}\right)^2 + K_2 \left(\frac{M_{x_1,\mathrm{approx}}(\mathbf x_i)}{M_\mathrm{s}} \right)^4 \right), \\
  L_\mathrm{ex}  &= \frac{\mu_0 V_\mathrm{dom}}{n} \sum_{i=1}^n in(\mathbf x_i) \left(\frac{C}{M_\mathrm{s}^2}\left[\left(\left.\nabla M_{x_1,\mathrm{approx}}(\mathbf x)\right|_{\mathbf x = \mathbf x_i} \right)^2 +\left(\left.\nabla M_{x_2,\mathrm{approx}}(\mathbf x)\right|_{\mathbf x = \mathbf x_i} \right)^2\right] \right).
\end{align}
Please note that the energies are scaled with the factor $\mu_0$. Here, we assumed that the anisotropy axis is parallel to the $x_2$ direction.

\section{Results}

We apply the Keras/Tensorflow wrapper SciAnn \cite{haghighat2021sciann} for implementing physics informed neural networks. We sample training points quasi-uniformly, applying the Sobol sequence as implemented in the Python library scikit-optimize \cite{head10scikit}.

\subsection{Magnetic field of a uniformly magnetized infinite prism}
\label{sec:magfield}
As test case, we pick a classical problem in micromagnetics \cite{gronefeld1989calculation}. We compute the magnetic field for a uniformly magnetized particle with rectangular cross section using the collocation based physics informed neural networks and the deep Ritz method. We consider a magnetic prism infinitely extended in the $x_3$ direction. It is uniformly magnetized in $x_2$ direction with the magnetization $M_\mathrm{s}$. The lower left corner and upper right corner of the rectangle are $(x_{1,\mathrm{min}},x_{2,\mathrm{min}})$ and $(x_{1,\mathrm{max}},x_{2,\mathrm{max}})$, respectively. The magnetic induction $\mathbf B = \mu_0 \mathbf H + \mu_0 \mathbf M$ inside the magnet is \cite{gronefeld1989calculation}
\begin{align}
\label{eq:groenefeld1}
B_{x_1} &= -\frac{\mu_0 M_\mathrm{s}}{4\pi}\ln \frac{\left( \left(x_1-x_{1,\mathrm{min}} \right)^2 + \left(x_2-x_{2,\mathrm{min}} \right)^2 \right)\left( \left(x_1-x_{1,\mathrm{max}} \right)^2 + \left(x_2-x_{2,\mathrm{max}} \right)^2 \right)}{\left( \left(x_1-x_{1,\mathrm{min}} \right)^2 + \left(x_2-x_{2,\mathrm{max}} \right)^2 \right)\left( \left(x_1-x_{1,\mathrm{max}} \right)^2 + \left(x_2-x_{2,\mathrm{min}} \right)^2 \right)},\\
\label{eq:groenefeld2}
B_{x_2} &= -\frac{\mu_0 M_\mathrm{s}}{4\pi} \left(\arctan \frac{x_1-x_{1,\mathrm{min}}}{x_2-x_{2,\mathrm{min}}}-\arctan \frac{x_1-x_{1,\mathrm{min}}}{x_2-x_{2,\mathrm{max}}} \right.\\ \nonumber
&\left.\;\;\;\;\;\;\;\;\;\;\,-\arctan \frac{x_1-x_{1,\mathrm{max}}}{x_2-x_{2,\mathrm{min}}}+\arctan \frac{x_1-x_{1,\mathrm{max}}}{x_2-x_{2,\mathrm{max}}}\right)+ \mu_0 M_{\mathrm s}.
\end{align}

We set the corners $(x_{1,\mathrm{min}},x_{2,\mathrm{min}})$ and $(x_{1,\mathrm{max}},x_{2,\mathrm{max}})$ to  $(-0.5,-0.5)$ and $(0.5,0.5)$, respectively. 

\begin{figure}[!tb]
	\centering
	\small{collocation} \hspace{3cm} \small{Ritz} \hspace{3cm} \small{ground truth~~~~~~~~~~~~~~~~~~} \\
	\includegraphics[scale=0.63,trim={1.2cm 0.3cm 9.6cm 0.8cm},clip]{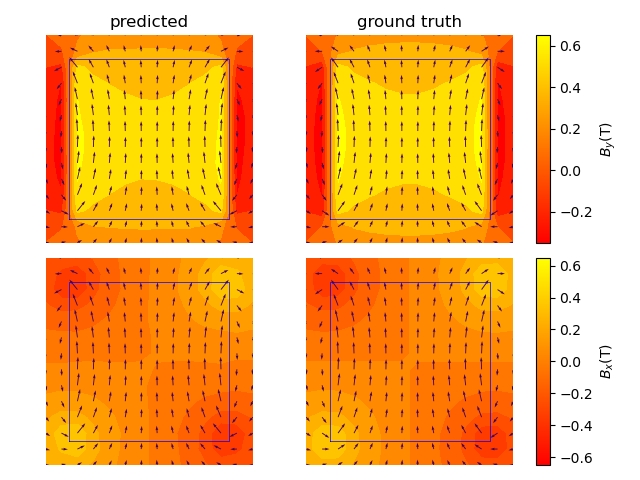}
	\includegraphics[scale=0.63,trim={0 0.3cm 0.5cm 0.8cm},clip]{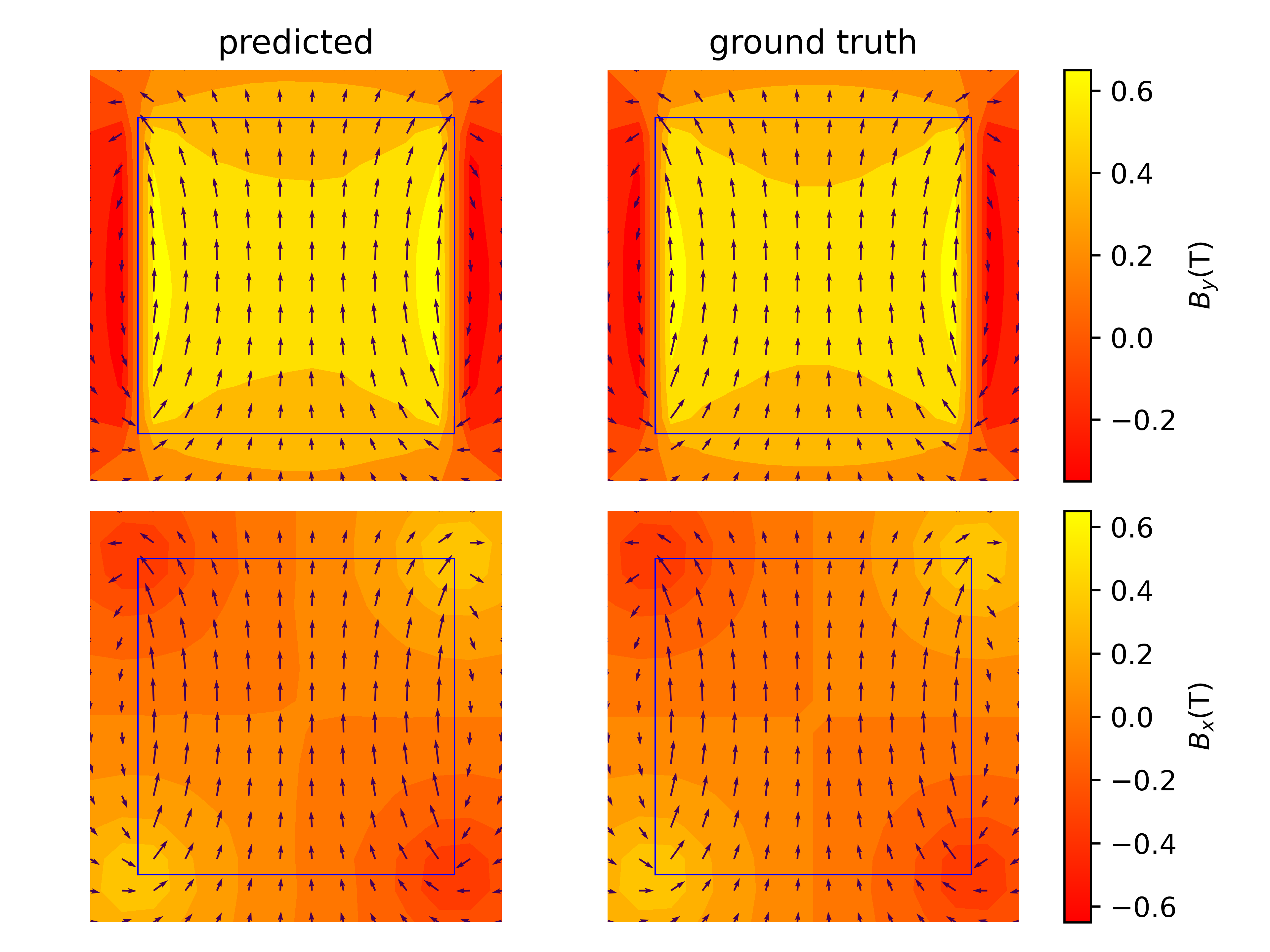}
	\caption{\label{fig:field}Demagnetizing field for the uniform magnetic particle with square cross-section. Left: Collocation method. Center: Ritz method. Right: Analytical solution.}
\end{figure}

\begin{table}[!htb]
	\renewcommand{\arraystretch}{1.2}
	\begin{center}
		\caption{Comparison of physics informed neural networks based on the collocation method and the Ritz method for solving a magnetostatic problem. In last line give the number of passes through the entire training set (epochs). At the given number of epochs the learning rate reached its minimum value.}
		\label{tab:magnetostatic}
		\begin{tabular}{l l c c} 
			\hline \hline
			quantity &	equation & collocation method & Ritz method \\ \hline
			magnetostatic energy & $-(\mu_0/2) \int_{V^{\mathrm{(in)}}} \mathbf H \cdot \mathbf M \mathrm{d}^2x$ & $0.255 \mu_0 M_\mathrm{s}^2 V^{\mathrm{(in)}}$ & $0.248 \mu_0 M_\mathrm{s}^2 V^{\mathrm{(in)}}$\\
			mean absolute error & $(1/m)\sum_i^m\left|\mathbf B_\mathrm{true}(\mathbf x_i) - \mathbf B_\mathrm{approx}(\mathbf x_i)\right|$ & 0.014~T & 0.011~T \\
			epochs & when learning rate $< 10^{-8}$ & 604 & 527 \\ \hline \hline
		\end{tabular}
	\end{center}
\end{table}

The magnetostatic energy density the infinite prism \cite{schrefl1994two} is $\mu_0 M_\mathrm{s}^2/4$. This result can be derived from the theorem that the magnetostatic self-energy of a uniformly magnetized particle of arbitrary shape is equivalent to that of suitably chosen ellipsoid of the same volume \cite{brown1957effect} and the demagnetizing factor $D = 1/2$ of an infinite prism with square cross-section \cite{aharoni1998demagnetizing}. For our simulations, we set $\mu_0 M_\mathrm{s}=1$~T. 

The problem domain extends outside the magnetic region. It is a square ranging from (-5,-5) to (5,5). In other words, we truncate the problem domain for the magnetic vector potential at $x_1 = \pm5$ and $x_2 = \pm5$. All networks have identical layout with 8 hidden layers with 32 neurons each. We used $\tanh$ as activation function. 

The total number of sampling points is $N = 2^{24}$.  We split the training set into $2^6$ batches of size $n =2^{18}$. For the indicator $in$ function we used
\begin{align}
 in(x_1,x_2) &= \left(\frac{\mathrm{sgn}(x_1+0.5)}{2}-\frac{\mathrm{sgn}(x_1-0.5)}{2}\right)\left(\frac{\mathrm{sgn}(x_2+0.5)}{2}-\frac{\mathrm{sgn}(x_2-0.5)}{2}\right). 
\end{align}
The indicator functions $bnd$ and $inf$ were set to 1 when the distance of a point to the respective boundary was less than $10^{-3}$.

For optimization, we applied the Adam method \cite{kingma2014adam} with an initial step size of $10^{-3}$. The number of complete passes through the training set (epochs) was 2000. However, this value is not reached owing to the following early stopping method. When there is no decrease in the  loss function for ten epochs the learning rate is reduced by $1/2$. Reducing the step size (learning rate) in a stochastic gradient descent method reduces the fluctuations in the loss.  The  different batches will give different gradients of the loss with respect to the weights which in turn causes random oscillations in the loss. A gentle decrease of the learning rate helps convergence \cite{hinton}.  Training is stopped if the learning rate reaches $\eta_{\mathrm{min}}=10^{-8}$.

Figure~\ref{fig:field} shows the computed magnetic flux density of a uniformly magnetized particle with rectangular cross-section. Virtually there is no difference between the results obtained from the collocation based physics informed neural network and deep Ritz method. A detailed comparison is given in Table~\ref{tab:magnetostatic}. The magnetostatic energy is computed by quasi-Monte-Carlo integration over the volume of the magnetic region $V^\mathrm{(in)}$. For integration $10^5$ points sampled with Sobol distribution are used. The same points are used to compute the mean absolute error.

\begin{figure}[!tb]
	\centering
	\includegraphics[scale=0.6]{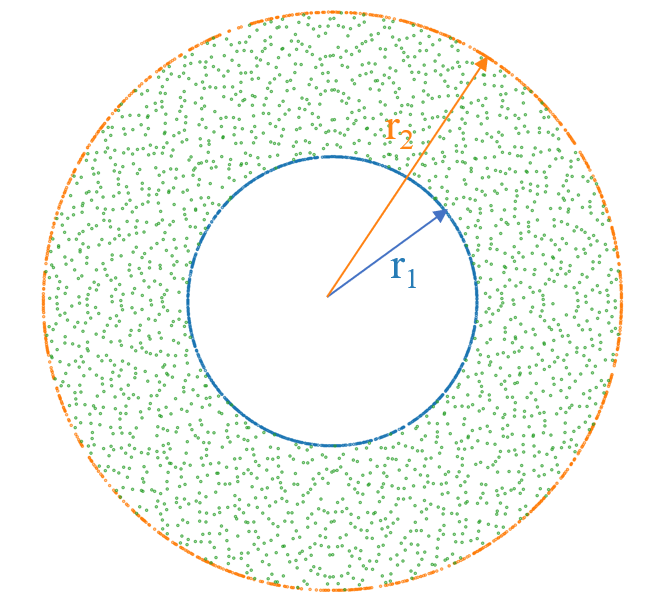}
	\caption{\label{fig:points}Points sampled for training of the physics informed neural network. The three set of points are sampled in the ring and at the inner and outer surface of the cylinder. For visibility the number of points was reduced as compared to the actual data used for training the network.}
\end{figure}

\subsection{Halbach cylinder}

We test the use of physics informed neural networks for the solution of a classical inverse magnetostatic problem. This can be achieved with permanent magnet flux sources with a well defined arrangement of permanent magnets \cite{halbach1985application}. We want to use the methodology outlined in section \ref{sec:inverse}, to  compute the orientation of the magnetization in a long magnetic cylinder that generates a uniform vertical field in a cylindrical cavity and is zero outside the magnetic system. 

The analytic solution is well-known \cite{halbach1980design}. Let $\theta$ denote the angle with the vertical axis. The inner radius and the outer radius of the magnetic hollow cylinder are $r_1$ and $r_2$. A uniform field is achieved when the magnetization $\mathbf{M}$ at any position $\theta$ in the magnet is rotated by the angle $2\theta$ with respect to the vertical axis. The uniform flux density in the cavity is  $B = \mu_0 M_s \ln(r_2/r_1)$. We set $\mu_0 M_s = 1$~T.

\begin{table}[!tb]
	\renewcommand{\arraystretch}{1.2}
	\begin{center}
		\caption{Minimal loss $L_{\mathrm{halbach}}$ achieved with different hyperparameters.  $w$ is the number of hidden layers, $l$ is the number of neurons per hidded layer, and $n$ is the batch size.
			For each set of hyperparameters  $L_{\mathrm{halbach}}$ and the number of passes through the training set (epochs) is given. The total number of training points was $N = 2^{20}$. The last column gives the mean absolute error, $(\mu_0/m)\sum_i^m\left|\mathbf{M}_\mathrm{true}(\mathbf x_i) - \mathbf{M}_\mathrm{approx}(\mathbf x_i)\right|$, computed with about $2.3 \times 10^5$ quasi-randomly sampled points. Owing to the inherent randomness of the algorithm slightly different values are obtained when the simulation is repeated for the same set of hyperparameters. The third column is a consecutive number for several runs with the same parameters.
		}
		\label{tab:halbach}
		\begin{tabular}{c l c r c l} 
			\hline \hline
			layout $h \times l$ &  batch size $n$ & run &  epochs & loss $L_{\mathrm{halbach}}$ & mean absolute error (T) \\ \hline
			$2\times 32$ & \hspace{0.65cm} $2^{10}$   &  1  &   2421   & $3.5 \times 10^{-4}$   &   \hspace{1.4cm} 0.089   \\
			$4\times 16$ & \hspace{0.65cm} $2^{14}$   &  1  &  10225   & $7.5 \times 10^{-4}$   &   \hspace{1.4cm} 0.13    \\
			$4\times 32$ & \hspace{0.65cm} $2^{8}$    &  1  &   2233   & $ 2.8 \times 10^{-5}$  &   \hspace{1.4cm}  0.021  \\
			$4\times 32$ & \hspace{0.65cm} $2^{10}$   &  1  &   1475   & $9.8 \times 10^{-5}$   &   \hspace{1.4cm} 0.055   \\
			$4\times 32$ & \hspace{0.65cm} $2^{14}$   &  1  &   8320   & $4.7 \times 10^{-4}$   &   \hspace{1.4cm} 0.1     \\
			$8\times 32$ & \hspace{0.65cm} $2^{8}$    &  1  &   1462   & $   1.0\times 10^{-8}$   &   \hspace{1.4cm} 0.0076   \\
			$8\times 32$ & \hspace{0.65cm} $2^{8}$    &  2  &   1292   & $   2.0\times 10^{-5}$   &   \hspace{1.4cm} 0.032   \\
			$8\times 32$ & \hspace{0.65cm} $2^{8}$    &  3  &   1848   & $   1.0\times 10^{-8}$   &   \hspace{1.4cm}  0.0059   \\
			$8\times 32$ & \hspace{0.65cm} $2^{8}$    &  4  &   1338   & $   9.6\times 10^{-9}$   &   \hspace{1.4cm} 0.011   \\
			$8\times 32$ & \hspace{0.65cm} $2^{8}$    &  5  &   2302   & $   1.0\times 10^{-8}$   &   \hspace{1.4cm} 0.0027   \\
			$8\times 32$ & \hspace{0.65cm} $2^{8}$    &  6  &   1623   & $   1.0\times 10^{-8}$   &   \hspace{1.4cm}  0.0027   \\
			$8\times 32$ & \hspace{0.65cm} $2^{8}$    &  7  &   2161   & $   1.0\times 10^{-8}$   &   \hspace{1.4cm} 0.0048   \\
			$8\times 32$ & \hspace{0.65cm} $2^{8}$    &  8  &   1800   &    2.3$ \times 10^{-5}$    &   \hspace{1.4cm} 0.032   \\
			$8\times 32$ & \hspace{0.65cm} $2^{10}$   &  1  &   1840   & $3.3 \times 10^{-5}$   &   \hspace{1.4cm} 0.028   \\
			$8\times 32$ & \hspace{0.65cm} $2^{14}$   &  1  &   5752   & $3.0 \times 10^{-4}$   &   \hspace{1.4cm} 0.1     \\
			\hline \hline
		\end{tabular}
	\end{center}
\end{table}

\begin{figure}[!tb]
	\centering
	\includegraphics[scale=0.37]{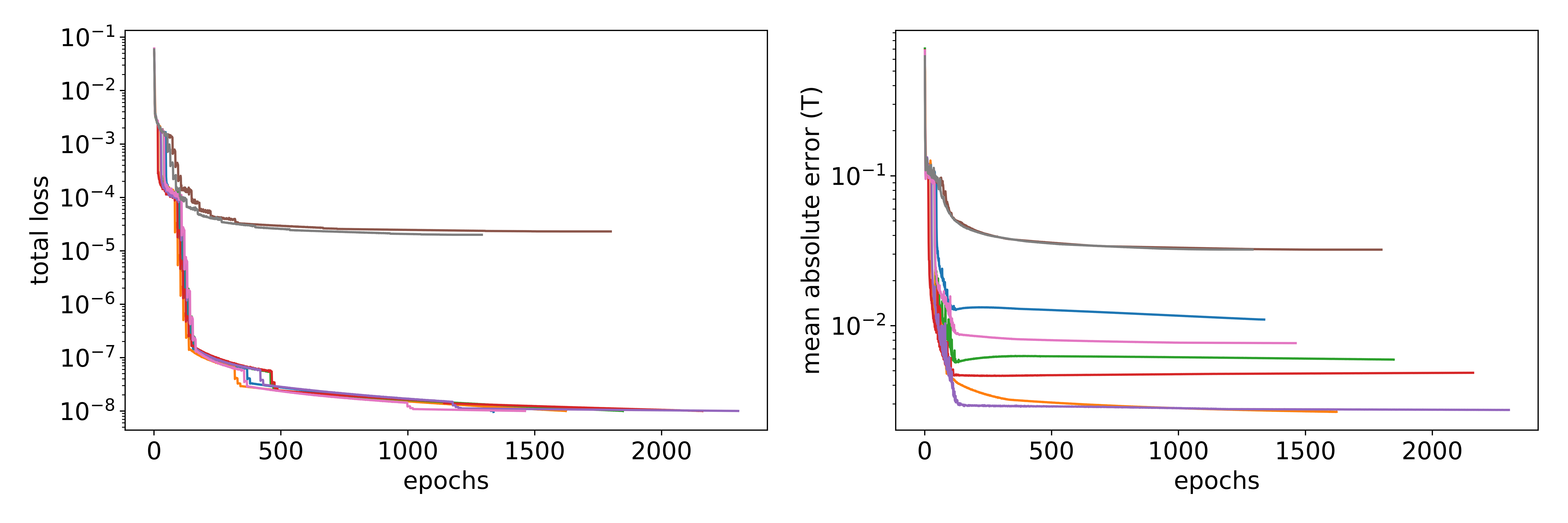}
	\caption{\label{fig:decrease}Decrease of the total loss (left hand side) and the mean absolute error (right hand side)  during training. $L_\mathrm{halbach}$ and $(\mu_0/m)\sum_i^m\left|\mathbf{M}_\mathrm{true}(\mathbf x_i) - \mathbf{M}_\mathrm{approx}(\mathbf x_i)\right|$ are plotted as a function of the number of epochs for eight different runs with the parameters $h \times l = 8 \times 32$ and $n = 2^8$.}
\end{figure}

\begin{figure}[!tb]
	\centering
	\includegraphics[scale=0.8,trim={0cm 0.8cm 0cm 0.8cm},clip]{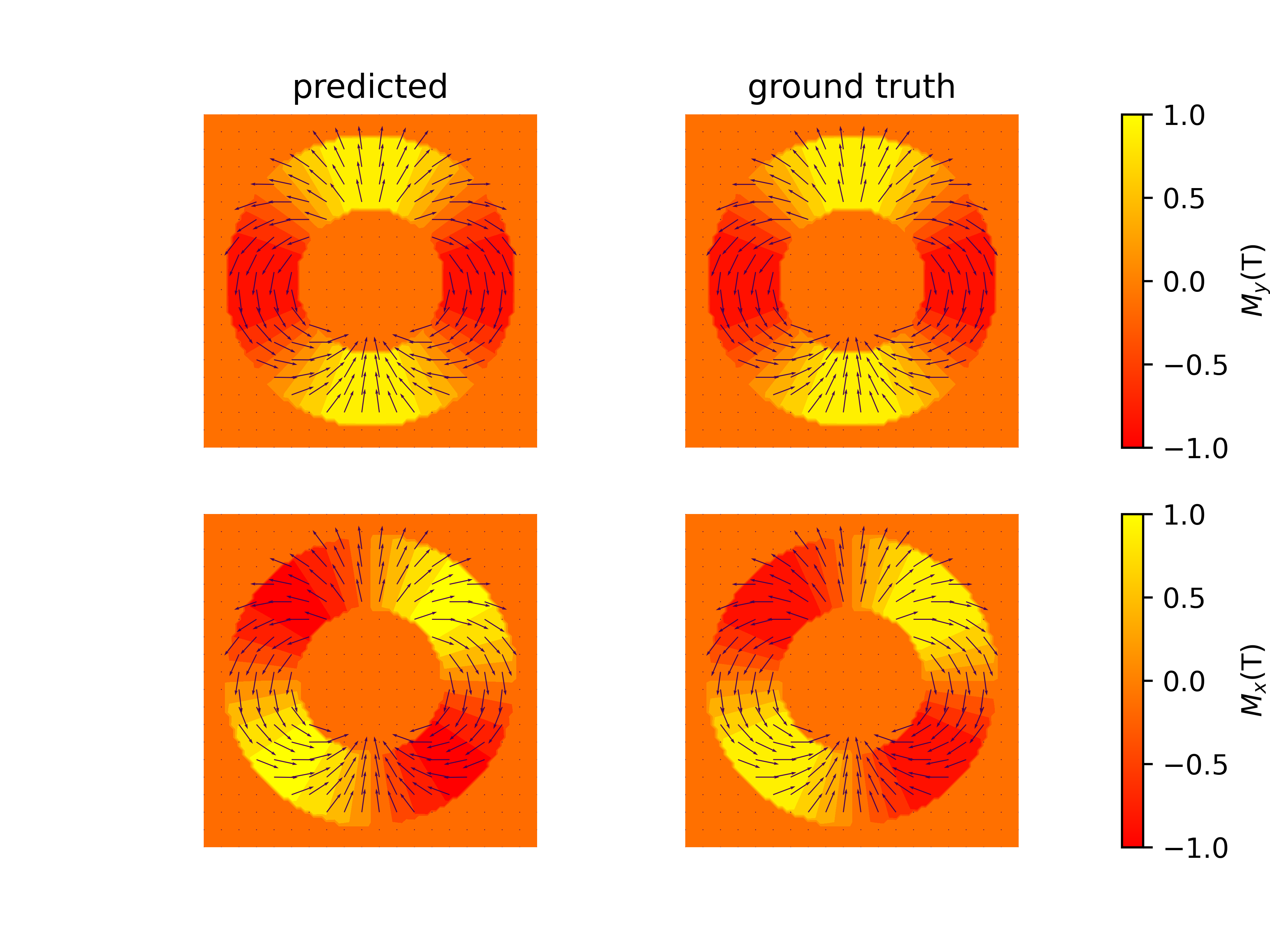}
	\caption{\label{fig:halbach}Comparison of the estimate from physics informed neural network and the analytical solution for a classical inverse magnetostatic problem. The computed magnetization for a Halbach cylinder is shown for $h \times l = 8 \times 32$ and $n = 2^8$  for $\mathrm{run}=1$ at $\mathrm{epoch}=1462$.}
	\vspace{0.2cm}
	\centering
	\includegraphics[scale=0.4]{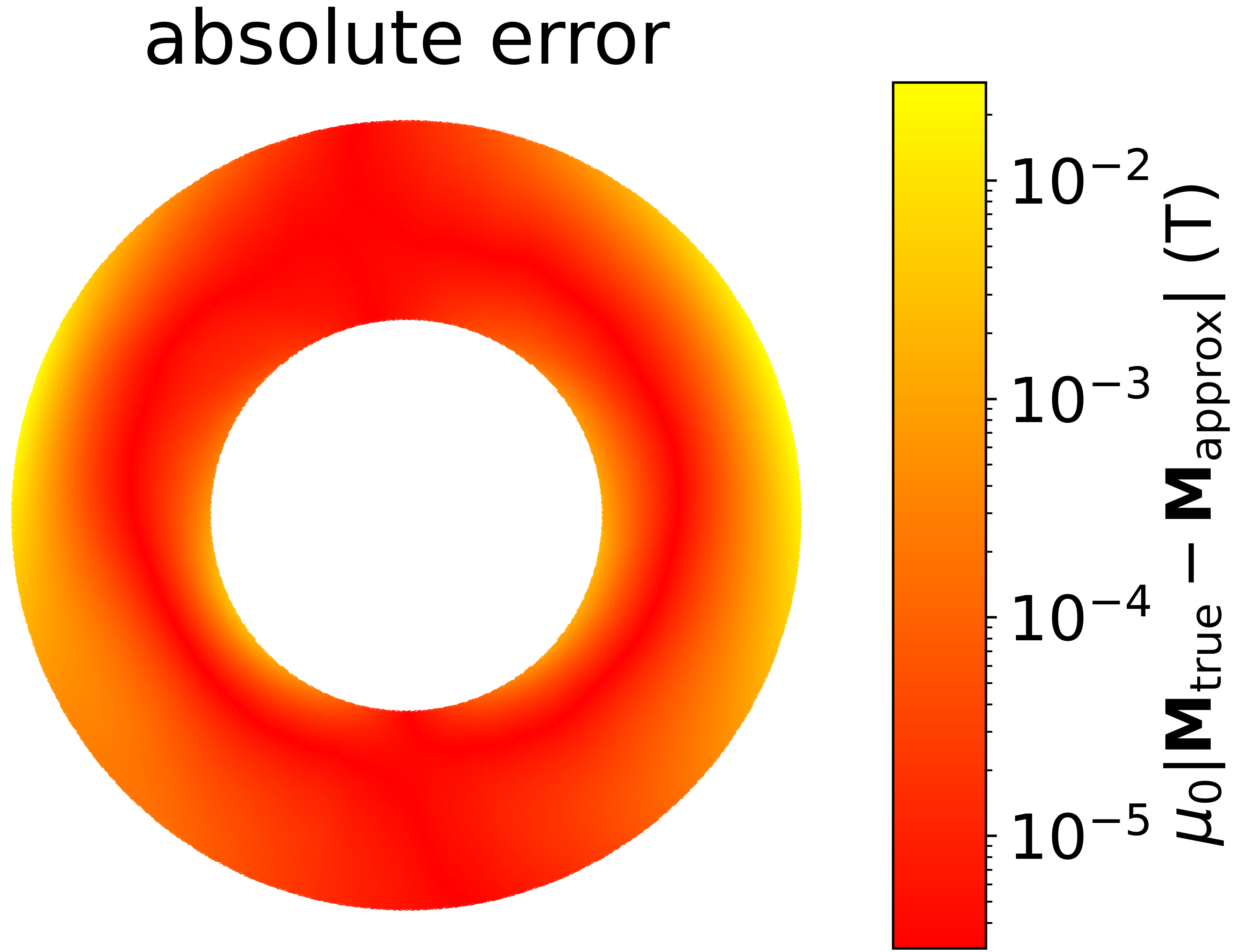}
	\caption{\label{fig:halbacherror} Absolute error $\mu_0\left|\mathbf{M}_\mathrm{true}(\mathbf x) - \mathbf{M}_\mathrm{approx}(\mathbf x)\right|$ between the magnetization of the true solution and the neural network
	approximation for the Halbach cylinder for $h \times l = 8 \times 32$ and $n = 2^8$ for $\mathrm{run}=1$ at $\mathrm{epoch}=1462$.}
\end{figure}

The inputs for the three networks $\mathcal{N}_{A}$, $\mathcal{N}_{M_{x_1}}$, and $\mathcal{N}_{M_{x_2}}$ were the position of the training points. The total number of points in the training set, $N$, is $N=2^{20}$. It consists of three distinct set of points. The points in the ring, points at the inner surface, and points at the outer surface. The indicator functions $in$, $bnd_1$, and $bnd_2$ were set accordingly. $N/2$ points are sampled quasi-randomly in an annulus with inner radius $r_1 = 1$ and outer radius $r_2 = 2$. We first create a Sobol sequence of quasi-random points in a square and then map the points \cite{assad2005circular} to the annulus. In addition, we create $N/4$ randomly sampled points on the outer surface and $N/4$ randomly sampled points at the inner surface. Figure~\ref{fig:points} shows the distribution of the training points. For visibility the number of points is lower than that actually used for training.

We tuned the hyperparameters of the network and for training with a manual search. Table~\ref{tab:halbach} compares the minimal value of the loss (\ref{eq:lossinv}) achieved with different set of hyperparameters. We modified the layout of the network $h\times l$  and the batch size $n$.  Here $h$ is the number of hidden layers that contain $l$ neuron each.  For optimization, we applied the Adam method \cite{kingma2014adam} with an initial step size of $10^{-3}$. We applied early stopping by the learing rate as described in section \ref{sec:magfield} with a minimum learning rate of $\eta_\mathrm{min} = 10^{-10}$. The number of maximum possible epochs was set to a very high number. The training was stoppend when the minimum learning rate was reached or the loss dropped below $10^{-8}$. Table \ref{tab:halbach} also lists the mean absolute error in the magnetization. The mean absolute error $(\mu_0/m)\sum_i^m\left|\mathbf{M}_\mathrm{true}(\mathbf x_i) - \mathbf{M}_\mathrm{approx}(\mathbf x_i)\right|$ was computed with about $2.3 \times 10^5$ quasi-randomly sampled points. The mean absolute error is not used for the selection of hyperparameters. For applications of physically informed neural networks for which the analytical solution is not known, only the total loss is available. 

The results listed in Table~\ref{tab:halbach} shows that reducing the batch size for fixed network layout reduces the total loss. This may be explained by the observations of Keskar and co-workers \cite{keskar2016large} who show that training with a large batch size finds minima which are much closer to the initial state than training with a smaller batch size. Methods with a smaller batch size explore the energy landscape and  move from the initial point towards minima that are located farther away. The minimum loss was obtained with $h \times l = 8 \times 32 $ and a batch size of $n = 2^8$. 

Repeated simulations with the same set of hyperparameters show different results \cite{reddi2018adaptive,shukla2020physics} owing to the inherent randomness of the stochastic gradient descent method. There are two reasons for randomness: The random initialization of the weights and randomly shuffled training sets. At the start of the algorithm the initial weights are randomly set by a truncated normal distribution. Each batch contains randomly picked points from the total training set. The gradients of the loss with respect to the weights will fluctuate from batch to batch when passing through the training data. Figure~\ref{fig:decrease} shows the total loss and the mean absolute error in the magnetization as a function of the number of full passes through the training set (epochs) for different runs with $h \times l = 8 \times 32 $ and $n = 2^8$. We observe a variance between the different runs. Loss and error drop rapidly at the beginning of the training for six out of eight runs. For two runs, the systems seems to be trapped in a bad local minimum. The use of adaptive activation functions \cite{jagtap2020adaptive,lu2021deepxde} can mitigate this problem. In practice, we can train the neural network several times with different initial seeds and use the weights that result in the lowest total loss for production runs.    

Figure~\ref{fig:halbach} compares the analytic solution and the estimate of the neural network for the Halbach cylinder computed with the hyperparameters $h \times l = 8 \times 32$, $n = 2^8$ for $\mathrm{run}=1$ at $\mathrm{epoch}=1462$. The vector plots show the magnetization in the hollow cylinder. The mean absolute error is 0.008~T.  Figure~\ref{fig:halbacherror} shows the distribution of the error within the ring.   

\subsection{Magnetization reversal of an infinite prism}
Here we apply the deep Ritz method for computing magnetization reversal of a magnetic particle as outlined in section \ref{sec:mumag}. The magnetization angle and the magnetic vector potential are approximated with two dense neural networks $\mathcal{N}_\varphi(x_1,x_2,\mathbf w_\varphi)$ and $\mathcal{N}_A(x_1,x_2,\mathbf w_A)$. The algorithmic framework for training neural networks is used to minimize the total Gibbs free energy for points along the demagnetization curve. 

Starting from a strong external field that saturates the particle, the repeated minimization of the Gibbs free energy gives the magnetic states along the demagnetization curve \cite{schrefl1994two}. The pretrained weights $\mathbf w_\varphi$ and $\mathbf w_A$ from the previous field step are used as initial weights for the successive training at the slightly decreased external field. In order to initialize the weights we used two steps. Firstly, we apply a strong external field and minimize the Zeeman energy by adjusting only $\mathbf w_\varphi$. Secondly, we keep $\mathbf w_\varphi$ fixed and minimize the magnetostatic energy by adjusting $\mathbf w_A$. For these two training steps, we apply the early stopping method discussed above. After this initial training, the networks $\mathcal{N}_\varphi$ and $\mathcal{N}_A$ give the magnetization angle and magnetic vector potential of the saturated state, respectively. In what follows we minimize the upper bound (\ref{eq:mumag}) for total energy for decreasing values of the external field by adjusting $\mathbf w_\varphi$ and $\mathbf w_A$ simultaneously. 

The demagnetization curve of a small hard magnetic particle has three characteristic branches. Initially the magnetization starts to deviate from the easy axis by reversible rotations. Especially near corners the magnetization rotates to minimize the magnetostatic energy \cite{gronefeld1989calculation}. The typical flower state is formed \cite{schabes1988magnetization}. During reversible processes the system follows a path of subsequent local minima \cite{kinderlehrer1994simulation}. At a critical value of the external field irreversible switching occurs. The system escapes from a saddle point towards the next minimum of the energy \cite{schabes1991micromagnetic}. Irreversible switching leads to the lower branch of the hysteresis loop. 

\begin{figure}[!t]
	\centering
	\textbf{undercritical state} ($\mu_0 H_\mathrm{ext} = -5.49$~T)\\
	\includegraphics[scale=0.6,trim={0cm 0cm 0cm 0.0cm},clip]{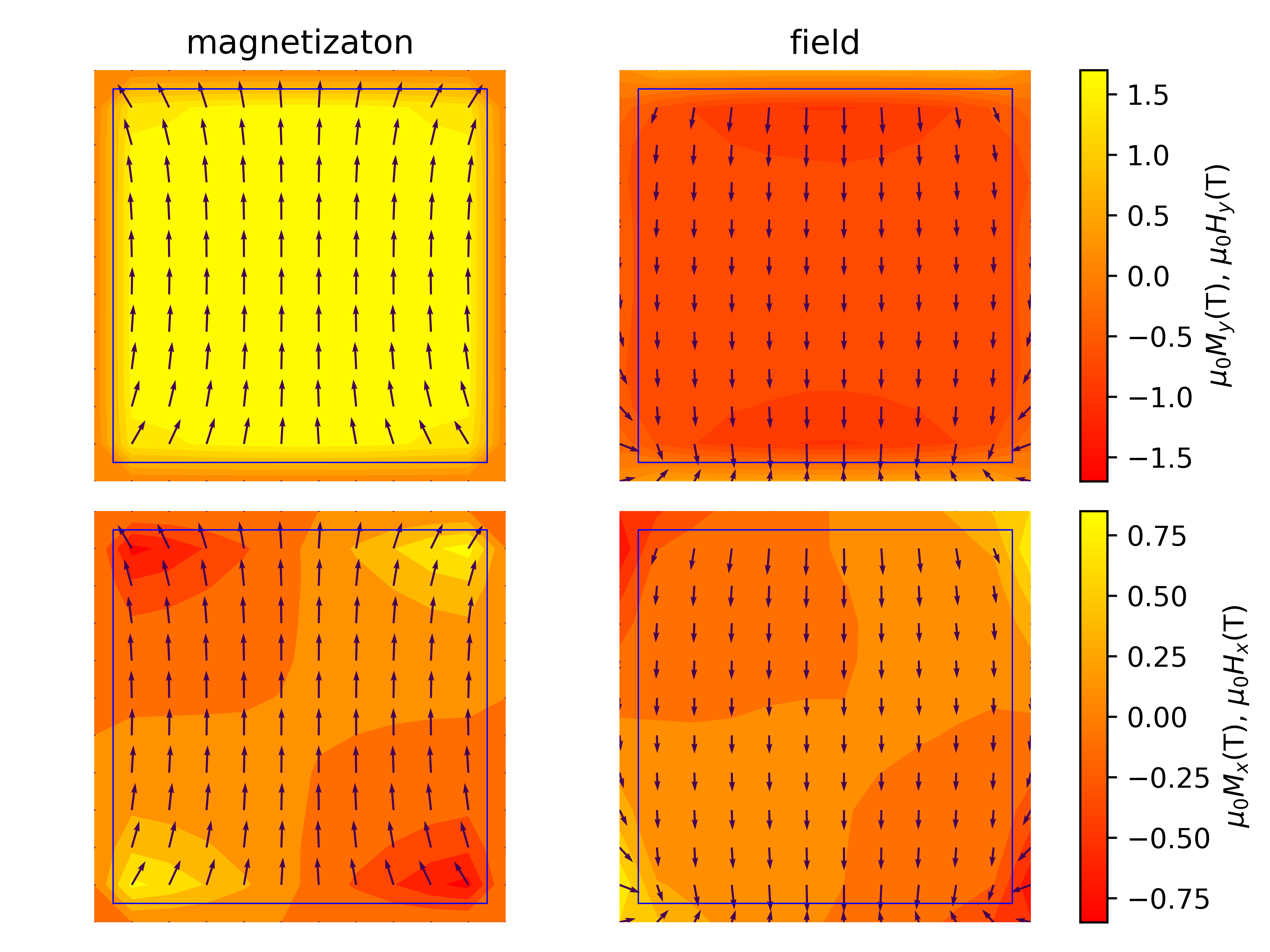} \\
	\textbf{switched} ($\mu_0 H_\mathrm{ext} = -5.5$~T) \\
	\includegraphics[scale=0.6,trim={0cm 0cm 0cm 0.0cm},clip]{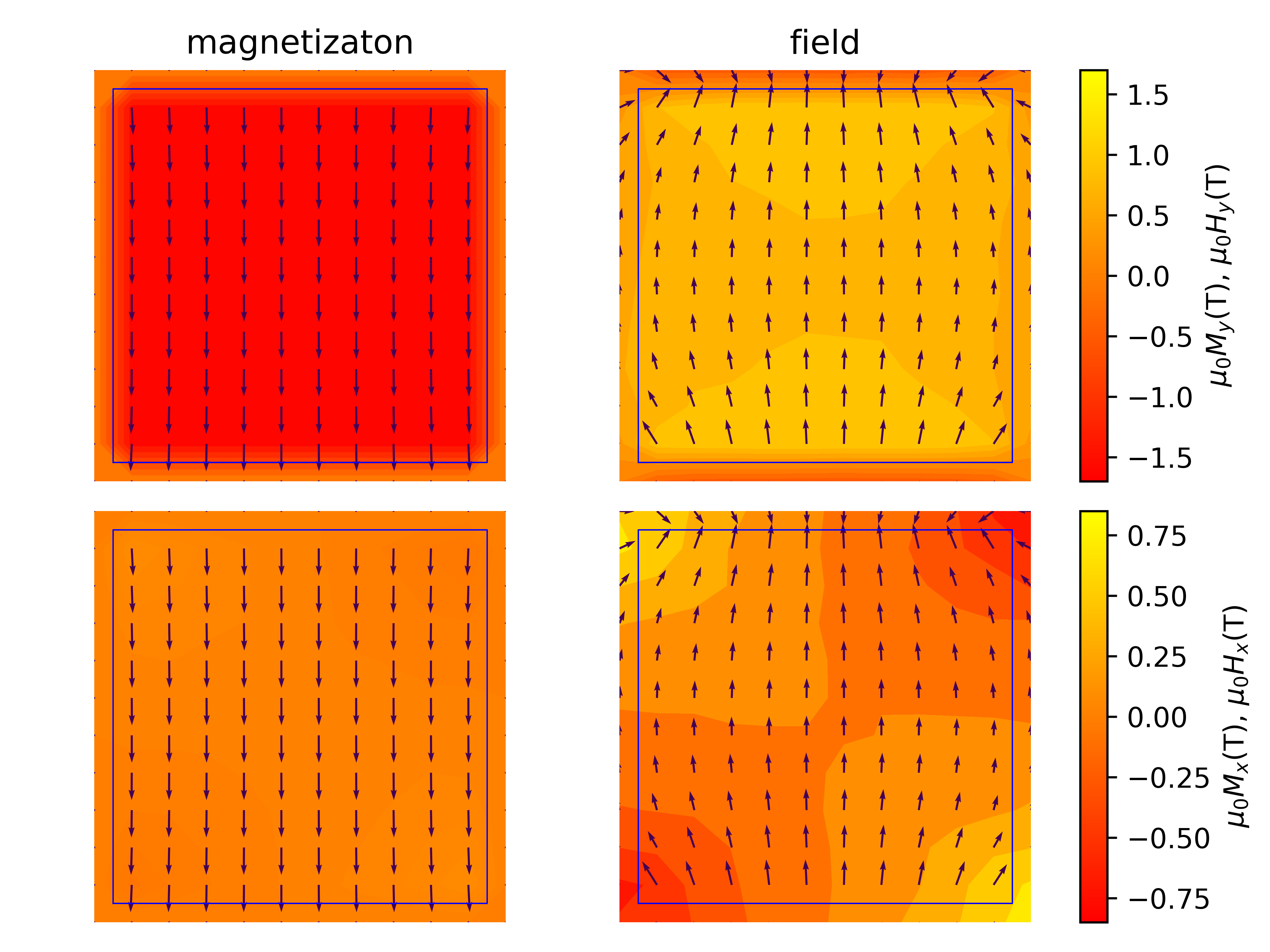}
	\caption{\label{fig:undercritical}Neural network approximation of the magnetization and the magnetic field before and after irreversible switching of a Nd$_2$Fe$_{14}$B particle with square cross-section. The easy axis lies vertically in the drawing plane. The particle is infinitely extended in the direction perpendicular to the drawing plane. }
\end{figure}

Optimizers used for training neural networks are designed to search for a deep local or global minimum of the loss function. In contrast, for computing hysteresis we want to follow a local minimum closely without escaping over a non-zero energy barrier. When computing the successive magnetic states along the demagnetization curve the energy should never increase for a fixed external field. On the other hand, once a saddle point is reached, we want to get out of the saddle point immediately. 

To meet the first requirement we modify the standard training method and apply an early stopping algorithm \cite{chollet2018deep}. We discard the current state, stop training, and move to the next field whenever an increase of the energy occurs during training. 

To get out of saddle points we use root mean square propagation (RMSprop) \cite{hinton} as optimizer. This is an adaptive gradient method which adapt the search direction by scaling the gradient. Let $g_i = \partial L / \partial w_i$ the component of the gradient of the loss function with respect to the weight $w_i$. The intuition behind RMSprop is to use an approximation of the sign of $g_i$ instead of $g_i$ as search direction. Then it is easy to move out regions with tiny gradients \cite{hinton}. The update rule of the $t$'s iteration is as follows:
\begin{align}
	\label{eq_rpmsprop1}
	MeanSquare_{\,i,t} &= \rho \, MeanSquare_{\,i,t-1} + (1-\rho)g^2_i, \\
	\label{eq_rpmsprop2}
	w_{i,t+1} &= w_{i,t} + \eta \frac{g_i}{\sqrt{MeanSquare_{\,i,t}} + \epsilon}.
\end{align} 
Here $\eta$ is the learning rate,  $\rho$ the discounting factor, and $t$ the iteration count. The total number of training points is split into batches. If $N$ is the number of training bounds and $n$ is the batch size, there are $N$/$n$ iterations to complete a full pass through the training set (epoch). The moving average of squared gradients in (\ref{eq_rpmsprop1}) smooths the mean square over adjacent batches. The parameter $\epsilon$ is regarded as a regularization term to avoid a large step when $\sqrt{MeanSquare_{\,i,t}}$ is close to zero.  However, it also controls the adaptivity level \cite{reddi2018adaptive}. Large values of $\epsilon$ reduce the influence of $MeanSquare_{\,i,t}$ and makes the algorithm more like stochastic gradient descent \cite{reddi2018adaptive,staib2019escaping}. Please note that ${g_i}/{\left(\sqrt{MeanSquare_{\,i,t}} + \epsilon\right)}$ is an approximation of the sign of the $i$-th component of the gradient, $\mathrm{sgn}(g_i) = {g_i}/{\sqrt{g_i^2}}$.   

Treating RMSprop as a preconditioned stochastic gradient descent method, Staib and co-workers \cite{staib2019escaping} found an optimal relation between the learning rate $\eta$ and the discounting factor $\rho$:
\begin{align}
	\rho_k = 1 - c \eta_k^{2/3}.
\end{align}
They also suggest using a decreasing step size. Therefore, we use $\eta_k = \eta_0/\sqrt{k}$. Here $k$ counts the full passes through the training set (epochs). We set the initial learning rate to $\eta_0 = 10^{-3}$.

We tuned the parameters $C$ and $\epsilon$ by comparing the computed switching field with the analytic result for a hard magnetic cylinder with circular cross-section. 

We simulate the reversal of Nd$_2$Fe$_{14}$B particles infinitely extended in a direction perpendicular to the particle's anisotropy axis \cite{schrefl1994two}. The intrinsic material parameters used for the simulations were $\mu_0 M_\mathrm{s} = 1.61$~T, $K_1 = 4.3$~MJ/m$^3$, $K_2 = 0.65$~MJ/m$^3$, and $C = 7.7$~pJ/m \cite{hock1988zuchtung}. The field step for the simulations was $\Delta \mu_0 H_\mathrm{ext} = 0.01$~T.  For a small particle with circular cross-section the demagnetizing field is uniform. For an external field applied parallel to the magneto-crystalline anisotropy axis the irreversible switching field is $2 K_1/(\mu_0 M_\mathrm{s})$ \cite{kronmuller1987angular}. This analytical value is $\mu_0 H_\mathrm{sw,true} = 6.71$~T for Nd$_2$Fe$_{14}$B. We computed the switching field using deep neural networks as outlined in section \ref{sec:mumag}. The total number of distinct training points was $N = 2^{24}$ and the batch size was $n =  2^{18}$. The dense neural networks had 8 hidden layers with 32 neurons each.  

We used simulations of the switching field for a cylinder with a diameter of 4.5~nm for tuning the hyperparemeters $c$ and $\epsilon$. We found that too small a value of $c$ caused premature switching: For $c < 1$ the magnet reverses at an external field  $|H_\mathrm{ext}| < H_\mathrm{sw,true}$ during the first pass through the training set, which indicates an unwanted escape from a local minimum.  For $c=10$ and $\epsilon = 10^{-7}$ the computed switching field, $\mu_0 H_\mathrm{sw,approx} = 6.74$~T. It is slightly higher than the analytical value, but the relative error is only 0.4 percent.  

We computed magnetization reversal for a Nd$_2$Fe$_{14}$B particle with a square cross-section. For an edge length of $L = 60$~nm the irreversible switching field computed with finite element micromagnetic simulations is $\mu_0 H_\mathrm{sw,fem} = 5.47$~T \cite{schrefl1994two}. The optimization of the neural network is a stochastic algorithm. Repeated simulations show slight fluctuations in the switching field approximated with the deep neural network. For $c=10$ and $\epsilon = 10^{-7}$, the switching fields from repeated runs varied in the range from $\mu_0 H_\mathrm{sw,approx} = 5.5$~T to $\mu_0 H_\mathrm{sw,approx} = 5.52$~T. The maximum relative error with respect to the analytic result was 0.9 percent. Figure~\ref{fig:undercritical} shows the computed magnetization and the demagnetizing field just before and after irreversible switching. The flower state is clearly seen in the undercritical state before switching.

\section{Conclusion}

We demonstrated the use of physics informed neural networks (PINNs) for magnetostatics, micromagnetics, and hysteresis computation. 

We used a deep neural network to approximate the magnetic vector potential. Training the neural network reduces the residuals of the static Maxwell equation at randomly sampled points in the problem domain, at its boundary and at material interfaces. To account for the interface condition of the magnetic flux density and the magnetic field at the surface of a magnetic material, separate networks for different regions were introduced \cite{niaki2020physics}. For the solution of inverse magnetostatic problems, we introduced additional neural networks that estimate the unknown magnetization. The loss function contains additional terms that penalize deviations from target conditions. The methodology was tested for the computation of the magnetization distribution in Halbach cylinders \cite{halbach1985application}.

Using Brown's upper bound for the magnetostatic energy, a deep Ritz method \cite{e2018deep} can be applied to solve magnetostatic field problems. The magnetostatic energy is the loss function for training the neural network. Adding the Zeeman energy, the ferromagnetic exchange energy density, and the magneto-crystalline anisotropy energy density, we built a micromagnetic solver that uses the algorithmic frame work of neural networks. Classical numerical schemes for micromagnetics are based on the very same energy functional \cite{asselin1986field,fredkin1987numerical,schrefl1994two}.

We believe that physics informed neural networks have great potential in computational magnetics. In particular, physics informed neural networks will come with some advantages: (1) There is no need for mesh generation. (2) Inverse problems may be solved effectively \cite{hennigh2020nvidia}. (3) A whole family of problems may be solved with a single neural network \cite{kovacs2021}.

\section*{Acknowledgment}
The financial support by the Austrian Federal Ministry for Digital and Economic  Affairs, the National Foundation for Research, Technology and Development and the Christian Doppler Research Association is gratefully acknowledged. 
L.E. and D.P. acknowledge support by the Austrian Science Fund (FWF) under grant No. P31140-N32 and grant No. F65, respectively.

\bibliographystyle{elsarticle-num-names}
\bibliography{references.bib}
\end{document}